# High-entropy perovskites, architectured by $s^0/d^0/d^{10}$ cations, as novel electrolytes for solid oxide fuel cells

Ho Truong Nam Hai[a,b], Rishad Kunafiev[c], Kwati Leonard[a,d], Kaveh Edalati[a,b,*]

[a] WPI, International Institute for Carbon Neutral Energy Research (WPI-I2CNER), Kyushu University, Fukuoka 819-0395, Japan

[b] Department of Automotive Science, Graduate School of Integrated Frontier Sciences, Kyushu University, Fukuoka 819-0395, Japan

[c] Graduate School of Engineering, Kyushu University, Fukuoka 819-0395, Japan

[d] Center for Energy System Design (CESD), International Institute for Carbon Neutral Energy Research (WPI-I2CNER), Kyushu University, Fukuoka 819-0395, Japan

Solid oxide fuel cells enable efficient conversion of hydrogen into electricity; however, the limited availability of materials for their cathode, anode, and electrolyte remains a concern. This study introduces three high-entropy oxide perovskites as novel electrolyte materials for fuel cells: $(Ba_{0.50}Sr_{0.50})(Ti_{0.33}Zr_{0.33}Hf_{0.33})O_3$ with $s^0/d^0$ cations, $(Ba_{0.50}Sr_{0.50})(Ga_{0.33}In_{0.33}Sn_{0.33})O_{3-\delta}$ with $s^0/d^{10}$ cations, and $(Ba_{0.50}Sr_{0.50})(Ti_{0.33}Zr_{0.33}Sn_{0.33})O_3$ with mixed $s^0/d^0/d^{10}$ cations. Through sequential sintering by high-pressure torsion (HPT) processing and calcination, these perovskites were synthesized and then printed with a thickness of about 6–10 µm on a Ni-$SrZr_{0.5}Ce_{0.4}Y_{0.1}O_{3-\delta}$ substrate as an anode and then coated with $Ba_{0.5}La_{0.5}CoO_{3-\delta}$ as a cathode. Electrochemical analysis and impedance spectroscopy show that $(Ba_{0.50}Sr_{0.50})(Ti_{0.33}Zr_{0.33}Sn_{0.33})O_3$ with $s^0/d^0/d^{10}$ cations exhibits the best performance with negligible current leakage and lowest ohmic resistance, while its maximum power density reaches 0.53 W.cm$^{-2}$ at 973 K. Complementary synchrotron X-ray absorption and photoelectron spectroscopy analyses indicate that the superior performance of $(Ba_{0.50}Sr_{0.50})(Ti_{0.33}Zr_{0.33}Sn_{0.33})O_3$ correlates with its heterogeneous electronic structure characterized by tailored unoccupied d-orbital states, and favorable local metal-oxygen bond lengths and extrinsic oxygen vacancies. This investigation demonstrates the significance of high-entropy perovskites with mixed $s^0/d^0/d^{10}$ cations as new rare-earth-metal-free ion-conducting electrolytes, particularly for protonic solid oxide fuel cells.

*Corresponding author (E-mail: kaveh.edalati@kyudai.jp; Phone: +81 92 802 6744)

## 1. Introduction

Countries around the world are reducing their use of fossil fuels. The consumption of green fuels is key to limiting environmental impacts, building a green society, and ensuring a renewable energy supply. Hydrogen is a clean energy carrier and an excellent alternative to fossil fuels [1]. In combustion or electrochemical processes, hydrogen produces significant energy and only water as a byproduct [2]. Recently, solid oxide fuel cells (SOFCs) that use hydrogen for energy storage and electricity generation have gained much attention [3-5]. However, most SOFCs, which are based on oxygen-ion conducting oxides, require high activation energies and operating temperatures, thereby making their applications and stability in electrochemical energy conversion more challenging [6]. One approach to reducing the operating temperature and enhancing the stability of SOFCs is to develop protonic ceramic fuel cells (PCFCs) [7].

A PCFC consists of three main components: the anode electrode, the cathode electrode, and the electrolyte [7]. PCFCs maintain a pure hydrogen supply at the anode and generate water at the cathode, thereby minimizing cathode oxidation and anode structural degradation [8]. In addition, PCFCs operate optimally at intermediate temperatures [9,10], thereby enhancing cell lifetime and reducing overall system cost [11]. The electrolyte layer is a high-density ceramic membrane that selectively conducts protons while blocking electrons and gases [12]. The electrolyte determines the position of water formation and largely controls both the performance and operating temperature of fuel cells [13]. To take advantage of proton-conducting electrolytes, perovskite oxides have been demonstrated to be effective for electrochemical applications in PCFCs [6,7]. Based on technical and structural requirements, perovskite electrolytes are designed to contain both large cations such as Ba, Sr, and La, and small cations such as Ce, Zr, Hf, Y, Yb, Ga, and Mg [6,7]. Such perovskites, including $SrCeO_3$ [14], $BaZrO_3$ [15], $BaCeO_3$ [16], $BaCe_{0.8}Y_{0.2}O_{3-\delta}$ [17], $SrZr_{0.9}Y_{0.1}O_{3-\delta}$ [18], $Ba_{1.05}Ce_{0.45}Zr_{0.1}Y_{0.1}Yb_{0.1}Pr_{0.1}Gd_{0.15}O_{3-\delta}$ [19], $BaZr_{0.44}Ce_{0.36}Y_{0.2}O_{3-\delta}$ [20], $LaGaO_3$ [21], and $LaAlO_{3-\delta}$ [21] have shown suitability as proton-conducting electrolyte materials. Although promising, these perovskite oxides still exhibit limitations, including sinterability issues during cell fabrication, limited chemical stability, low electrochemical performance, and reliance on rare-earth elements.

To improve electrochemical properties and optimize fuel cell performance, doping is widely used in perovskite electrolytes [22]. Motivated by the efficiency of doping in improving properties, the concept of high-entropy ceramics has recently been extended to various applications [23,24]. This concept involves incorporating at least five cation types into a single crystal lattice. The high-entropy strategy prioritizes incorporating beneficial elements into the main composition, thereby enhancing various properties, such as proton ionic conductivity [7] via the cocktail effect by increasing the proportion of favorable elements [23,24]. Moreover, high-entropy ceramics exhibit excellent structural stability owing to their low Gibbs free energy, arising from a high configurational entropy exceeding 1.5*R* (*R* is the gas constant) [23,24]. This stability issue can help maintain electrolyte durability in SOFCs [25,26]. Such structural stability allows the inclusion of elements with totally different natures, such as multiple anions [27], cations with different work functions [28], cations with high electronegativity mismatch [29], mixed $d^0/d^{10}$ cations [30], and mixed $d^0/d^{10}/s^0$ cations [31], giving the possibility to achieve unprecedented properties. The presence of a combination of elements in a lattice is also effective for tuning properties without the use of noble or rare-earth elements [32].

In this study, three $ABO_3$-type high-entropy perovskites (HEPs), $(Ba_{0.50}Sr_{0.50})(Ti_{0.33}Zr_{0.33}Hf_{0.33})O_3$, $(Ba_{0.50}Sr_{0.50})(Ti_{0.33}Zr_{0.33}Sn_{0.33})O_3$ and $(Ba_{0.50}Sr_{0.50})(Ga_{0.33}In_{0.33}Sn_{0.33})O_{3-\delta}$, are introduced as electrolytes for SOFC fabrication. The three perovskites have $s^0$ cations in the A site (Ba and Sr), while the B site contains only $d^0$ cations (Ti, Zr, and Hf) in the first oxide, mixed $d^0/d^{10}$ cations in the second oxide, and $d^{10}$ cations (Ga, In, and Sn) in the third oxide. The combination of $d^0$ metals (Ti, Zr, and Hf), whose d orbitals are empty, and $d^{10}$ metals (Ga, In, and Sn), whose d shells are filled, suppresses electron migration and current leakage because electrons cannot easily hop between these cations [22]. Furthermore, severe electronic configuration heterogeneity around A-site atoms (Ba and Sr) contributes to faster proton transport by generating structural heterogeneity/distortion at the atomic scale. The results show that $(Ba_{0.50}Sr_{0.50})(Ti_{0.33}Zr_{0.33}Sn_{0.33})O_3$ exhibits the best electrochemical performance in fuel cells, where $Ba_{0.5}La_{0.5}CoO_{3-\delta}$ is used as the cathode, and $Ni/SrZr_{0.5}Ce_{0.4}Y_{0.1}O_{3-\delta}$ is used as the anode. This study demonstrates the high potential of high-entropy perovskites with mixed $s^0/d^0/d^{10}$ cations as new rare-earth-metal-free electrolytes for SOFCs. Moreover, this study represents the first application of severe plastic deformation via the high-pressure torsion (HPT) [33] method to synthesize electrolyte materials for fuel cells.

## 2. Materials and methods

### *2.1. Reagent*

$(Ba_{0.5}La_{0.5})CoO_{3-\delta}$, $SrZr_{0.5}Ce_{0.4}Y_{0.1}O_{3-\delta}$ and NiO were prepared from Kusaka Rare Metal Products Co., Ltd., Japan, and Vogler, the Netherlands, respectively. SC-0708A Mariarim, as a dispersant, 3-hydroxy-2,2,4-trimethylpentyl isobutyrate (TMS) as a plasticizer, and TMS–ethylcellulose as a binder, were obtained from Nippon Oil & Fats Co., Ltd., Japan. High-purity $BaO_2$, $SrO_2$, $TiO_2$, $ZrO_2$, $HfO_2$, $Ga_2O_3$, $In_2O_3$, and $SnO_2$ were prepared from Kojundo, Japan or Sigma-Aldrich, USA.

### *2.2. Materials and Fabrication of Cells*

To fabricate fuel cells, three functional layers, including the electrolyte, anode electrode, and cathode electrode, were prepared. Three high-entropy perovskite $(Ba_{0.50}Sr_{0.50})(Ti_{0.33}Zr_{0.33}Hf_{0.33})O_3$, $(Ba_{0.50}Sr_{0.50})(Ti_{0.33}Zr_{0.33}Sn_{0.33})O_3$, and $(Ba_{0.50}Sr_{0.50})(Ga_{0.33}In_{0.33}Sn_{0.33})O_{3-\delta}$ were used as electrolyte layer in the fuel cells. These materials were synthesized by HPT followed by calcination. HPT is a powerful mechanical processing technique that simultaneously applies torsional strain (by rotating the sample between two anvils with a constant rotation speed) and compressive pressure (usually gigapascals), enabling nanometric-level mixing of precursor powders [34] and controlling solid-state reactions [35]. Subsequent calcination facilitates rapid atom-level mixing and stable phase formation. In addition, the HPT-based synthesis route offers the advantage of producing highly homogeneous and highly pure powders [36]. The synthesis was conducted in five steps. First, the binary oxides with stoichiometric masses were mixed in a mortar for about 30 min in the presence of ethanol. Second, the mixtures were compacted into discs with a 5 mm radius and a thickness of about 0.85 mm. Third, the compacted discs were treated via HPT under 6 GPa for 3 rotations at ambient temperature. Fourth, the sample after HPT was crushed and processed again under similar conditions. Fifth, the HPT-processed samples were ground and calcined at 1373 K for 24 h. Finally, the HPT and calcination processes were repeated under similar conditions to maximize compositional homogeneity at the atomic scale.

To prepare the electrolyte slurry, after perovskite synthesis, 1 g of each powder and 5 g of ethanol were placed in a zirconia pot and milled with zirconia balls (3 mm in diameter, 20 g) for 1 h at 500 rpm. The milled powders were dried and sieved through a 150 μm mesh to achieve a uniform particle size. Each pre-treated powder was then combined with 20 mg of dispersant, 0.258 g of TMS solvent as plasticizer, and 0.6 g of TMS-ethylcellulose as binder. The mixture was milled in a zirconia vial using balls for 1 h at 300 rpm to obtain a homogeneous electrolyte slurry. To make the anode electrode, a single-layer substrate was prepared from $SrZr_{0.5}Ce_{0.4}Y_{0.1}O_{3-\delta}$ and NiO powders at a mass ratio of 40:60, with a total mass of 20 g. The powders were first mechanically blended in a zirconia mortar, then transferred to a zirconia vial, and milled for 1 h at 300 rpm using 3 g of ethanol and 4 zirconia balls (15 mm diameter). The milled powder was dried under an infrared lamp at 175 W for 2 h, then sieved through a 150 μm mesh to achieve a uniform particle size. In the next step, 1.3 g of powder was compressed into a 20 mm diameter disc using the solid pelletizing method and subjected to 10 min of cold isostatic pressing at 250 MPa to obtain homogeneous, high-density green compacts. The compacts were calcined in a muffle furnace according to the following thermal program: ramping from room temperature to 1073 K over 8 h, holding for 5 h, and cooling down to ambient temperature over 4 h. After calcination, the compacts were polished to form a substrate with a thickness of 1.2 mm. Fine $(Ba_{0.5}La_{0.5})CoO_{3-\delta}$ powder, whose structure is shown in Fig. S1, was mixed with a 6 wt% ethylcellulose–terpineol binder to prepare the cathode slurry. Mixing was performed using a ball-milling setup identical to that used for anode preparation. The resulting slurry was deposited on the electrolyte surface by screen printing.

After obtaining the electrolyte slurry, cathode slurry, and anode substrate, the cells were fabricated using a four-step procedure. First, the electrolyte slurry was printed onto the $Ni/SrZr_{0.5}Ce_{0.4}Y_{0.1}O_{3-\delta}$ single-layer substrate (the structure of which is shown in Fig. S2) using a screen-printing process with a 15.5-mm mesh. Two printing passes were applied for the $(Ba_{0.50}Sr_{0.50})(Ti_{0.33}Zr_{0.33}Hf_{0.33})O_3$ and $(Ba_{0.50}Sr_{0.50})(Ti_{0.33}Zr_{0.33}Sn_{0.33})O_3$ electrolytes, while six printing passes were used for the $(Ba_{0.50}Sr_{0.50})(Ga_{0.33}In_{0.33}Sn_{0.33})O_{3-\delta}$ electrolyte. Second, the printed compacts were sintered for 10 h at 1673 K for $(Ba_{0.50}Sr_{0.50})(Ti_{0.33}Zr_{0.33}Hf_{0.33})O_3$ and 1653 K for $(Ba_{0.50}Sr_{0.50})(Ti_{0.33}Zr_{0.33}Sn_{0.33})O_3$ to obtain dense half-cells. For $(Ba_{0.50}Sr_{0.50})(Ga_{0.33}In_{0.33}Sn_{0.33})O_{3-\delta}$, significant compositional volatilization and the formation of sintering-induced voids were observed at 1673 K for 5-10 h (Fig. S3 and S4). Therefore, the sintering conditions for fabricating the electrolyte layer of $(Ba_{0.50}Sr_{0.50})(Ga_{0.33}In_{0.33}Sn_{0.33})O_3$ were optimized to 1573 K for 5 h. Third, the half-cells were coated four times with cathode slurry onto the electrolyte layer via screen printing to form fuel cells. Finally, silver-palladium paste as a current collector and platinum wire sets as current leads were attached to the cells.

### *2.3. Materials characterization*

The crystal structures of $(Ba_{0.50}Sr_{0.50})(Ti_{0.33}Zr_{0.33}Hf_{0.33})O_3$, $(Ba_{0.50}Sr_{0.50})(Ti_{0.33}Zr_{0.33}Sn_{0.33})O_3$, and $(Ba_{0.50}Sr_{0.50})(Ga_{0.33}In_{0.33}Sn_{0.33})O_{3-\delta}$ were determined by X-ray diffraction (XRD) using Cu K$\alpha$ radiation at 45 kV and 200 mA. XRD spectra were refined utilizing the Rietveld method and corresponding reliability factors $R_p$ (profile residual), $R_{wp}$ (weighted profile residual), and $\chi 2$ (chi-squared value). Scanning electron microscopy (SEM) provided cross-sectional micrographs of single cells at 5 keV, while a higher accelerating voltage of 15 keV was used to assess perovskite morphology. The nanoscale structures of perovskites were characterized using high-resolution transmission electron microscopy (TEM) at 200 keV. Elemental distribution in the HEPs was examined by energy-dispersive X-ray spectroscopy (EDS) in SEM and in a scanning transmission electron microscope (STEM). SEM samples were prepared by dispersing powders on carbon tape, and TEM/STEM samples by pouring the powders on carbon-coated copper grids. Local electronic configuration and atomic environment were explored by X-ray absorption spectroscopy (XAS), analyzing X-ray absorption near-edge structure (XANES) and extended X-ray absorption fine structure (EXAFS) regions. Samples for XAS were prepared by mixing perovskites with BN and pressing them into 10 mm diameter discs. XAS was conducted in the synchrotron facility of Saga Light Source using beamline BL11. X-ray photoelectron spectroscopy (XPS) with Al Kα radiation was used to characterize the oxygen states and associated vacancies. Electron spin resonance (ESR) at $9.4688 \times 10^9$ Hz was used to examine the presence of unpaired electrons corresponding to vacancies.

### *2.4. Electrochemical measurements*

Fuel cells were sealed in a Probostat rig (NorECS, Oslo, Norway) using Pyrex glass. To check for leakage based on open-circuit voltage characteristics, high-purity $H_2$ was humidified with saturated water vapor at 300 K ($P_{H_2O}$ = 1.9 kPa) and then supplied to the anode as the sweep gas at a flow rate of 100 mL.min$^{-1}$. At the same time, wet air was fed to the cathode at the same flow rate. Electrochemical impedance spectroscopy (EIS) of the fabricated cells was evaluated under dry and wet conditions of $H_2$ and air, and at various temperatures. The EIS spectra were measured at 773-973 K under open-circuit voltage using frequencies of 1 MHz to 0.1 Hz, with a 10 mV AC amplitude, utilizing a potentiostat (SP–300/240 Module, Bio-Logic Science Instrument, France). The ZView software was used to analyse and fit the impedance spectra. Moreover, the current–voltage and polarization curves at different temperatures were measured to further clarify the electrochemical properties of cells.

## 3. Results

### *3.1. Overall characterizations*

To determine the configurational entropy of the electrolyte materials, the configurational entropy at the A site ($\Delta S_{\mathrm{A,config}}$), B site ($\Delta S_{\mathrm{B,config}}$), and overall composition ($\Delta S_{\mathrm{config}}$) was calculated using Eq. 1. In addition, the tolerance factor ($t$) and octahedral factor ($\mu$) were evaluated using Eq. 2 and 3, respectively, in order to assess the geometric compatibility and structural feasibility of these newly developed perovskites [32].

$$\Delta S_{\text{config}} = \Delta S_{\text{A,config}} + \Delta S_{\text{B,config}} = -R \sum_{i=1}^{N_A} x_{i,\text{A}} ln x_{i,\text{A}} + -R \sum_{j=1}^{N_B} x_{j,\text{B}} ln x_{j,\text{B}} \quad (1)$$

$$t = \frac{\bar{r}_A + r_X}{\sqrt{2}(\bar{r}_B + r_X)} \quad (2)$$

$$\mu = \frac{\sum_{j=1}^{M} r_{j,B}}{3 r_X} \quad (3)$$

where $R$ is the gas constant. $N_A$ and $N_B$ are the numbers of distinct principal elements occupying the A and B sites, respectively; $x_{i,A}$ is the mole fraction of the $i$th element on the A site, $x_{j,B}$ is the mole fraction of the $j$th element on the B site; $\bar{r}_A$, $\bar{r}_B$ and $r_X$ denote the average ionic radius of A-site cations, B-site cations, and anion, respectively; $r_{j,B}$ represents the ionic radius of the $j$th B-site element; and $M$ represents the total number of B-site elements.

As shown in Table 1, the total configurational entropy for all three materials is the same (1.79$R$). This value exceeds 1.5$R$, indicating that these materials can be classified as high-entropy oxides [23]. In addition, Table 1 presents the ionic radii of the A-site ions (Ba, and Sr) with a coordination number of 12 and the B-site ions with a coordination number of 6. Based on these values, the calculated tolerance factors range from 0.9 to 1.0, while the octahedral factors range from 0.4 to 0.8 for 3 HEPs. These values geometrically ensure the arrangement of the constituent elements within an ideal cubic structure [32].

Table 1. Theoretical values of configurational entropy and geometric characteristics of three high-entropy perovskites $(Ba_{0.50}Sr_{0.50})(Ti_{0.33}Zr_{0.33}Hf_{0.33})O_3$, $(Ba_{0.50}Sr_{0.50})(Ti_{0.33}Zr_{0.33}Sn_{0.33})O_3$, and $(Ba_{0.50}Sr_{0.50})(Ga_{0.33}In_{0.33}Sn_{0.33})O_{3-\delta}$. XIIc.s. and VIc.s. are coordination numbers 12 and 6.

| **Composition** | **Entropy configuration** | **A-site ions (XIIc.s.)** | | **B-site ions (VIc.s.)** | | | | | | **X ion** | **Tolerance factor** | **Octahedral factor** |
|---|---|---|---|---|---|---|---|---|---|---|---|---|
| | | **$Ba^{2+}$** | **$Sr^{2+}$** | **$Ti^{4+}$** | **$Zr^{4+}$** | **$Hf^{4+}$** | **$Ga^{3+}$** | **$In^{3+}$** | **$Sn^{4+}$** | **$O^{2-}$** | | |
| $(Ba_{0.50}Sr_{0.50})(Ti_{0.33}Zr_{0.33}Hf_{0.33})O_3$ | 1.79 $R$ | 1.60 | 1.44 | 0.74 | 0.86 | 0.85 | - | - | - | 1.26 | 0.946 | 0.649 |
| $(Ba_{0.50}Sr_{0.50})(Ti_{0.33}Zr_{0.33}Sn_{0.33})O_3$ | 1.79 $R$ | 1.60 | 1.44 | 0.74 | 0.86 | - | - | - | 0.83 | 1.26 | 0.949 | 0.644 |
| $(Ba_{0.50}Sr_{0.50})(Ga_{0.33}In_{0.33}Sn_{0.33})O_{3-\delta}$ | 1.79 $R$ | 1.60 | 1.44 | - | - | - | 0.76 | 0.94 | 0.83 | 1.26 | 0.935 | 0.669 |

The crystal structures of the electrolyte materials were first determined using XRD. As described in Fig. 1 for (a) $(Ba_{0.50}Sr_{0.50})(Ti_{0.33}Zr_{0.33}Hf_{0.33})O_3$, (c) $(Ba_{0.50}Sr_{0.50})(Ti_{0.33}Zr_{0.33}Sn_{0.33})O_3$, and (e) $(Ba_{0.50}Sr_{0.50})(Ga_{0.33}In_{0.33}Sn_{0.33})O_{3-\delta}$, XRD spectra and Rietveld refinement confirm that the three materials have a cubic phase with a $Pm\overline{3}m$ space group. This crystal structure corresponds to an ideal perovskite structure with an $a^0a^0a^0$ tilting system and an average B–O–B bond angle of 180°. Therefore, no significant long-range octahedral tilting is expected in their crystal structures. However, due to the presence of multiple cations with different ionic radii occupying the B-site, local lattice distortions may still exist at the short-range scale, although these distortions are not significant enough to induce detectable symmetry lowering from the $Pm\overline{3}m$ cubic structure. Among the three perovskites, $(Ba_{0.50}Sr_{0.50})(Ti_{0.33}Zr_{0.33}Sn_{0.33})O_3$ exhibits the smallest lattice parameter, as presented in Table 2. Nanometric observation through high-resolution TEM imaging and fast Fourier transform (FFT), in Fig. 1 for (b) $(Ba_{0.50}Sr_{0.50})(Ti_{0.33}Zr_{0.33}Hf_{0.33})O_3$, (d) $(Ba_{0.50}Sr_{0.50})(Ti_{0.33}Zr_{0.33}Sn_{0.33})O_3$, and (f) $(Ba_{0.50}Sr_{0.50})(Ga_{0.33}In_{0.33}Sn_{0.33})O_{3-\delta}$, also confirms the presence of the cubic phase in their microstructure. These structural analyses confirm the successful synthesis of cubic perovskites, which have high potential for use in fuel cells [37,38].

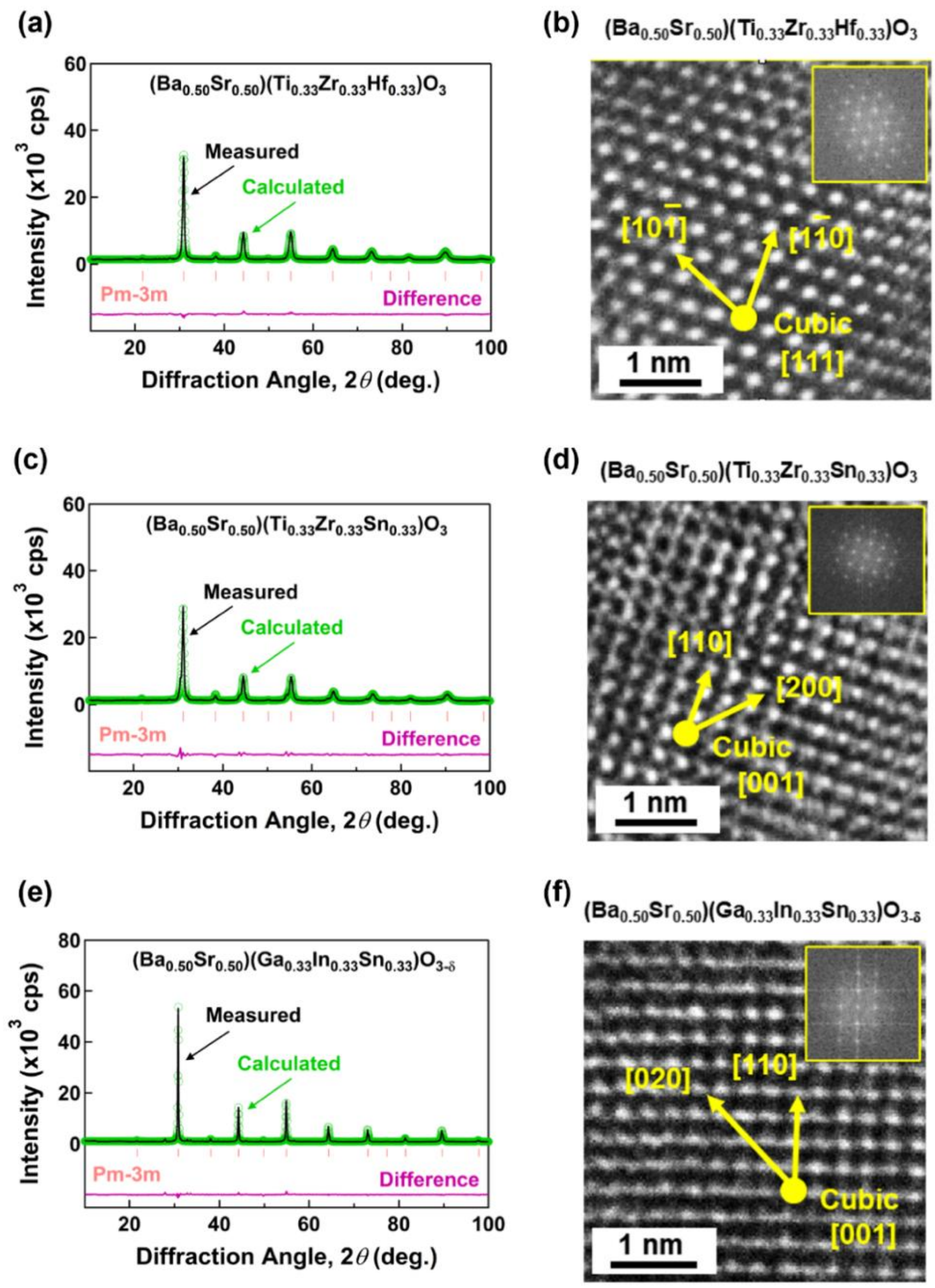

Fig. 1. Cubic perovskite phase of three high-entropy oxides. XRD pprofiles with relevant Rietveld refinement (left side) and high-resolution TEM micrographs with FFT diffractograms (right side) for (a, b) $(Ba_{0.50}Sr_{0.50})(Ti_{0.33}Zr_{0.33}Hf_{0.33})O_3$, (c, d) $(Ba_{0.50}Sr_{0.50})(Ti_{0.33}Zr_{0.33}Sn_{0.33})O_3$, and (e, f) $(Ba_{0.50}Sr_{0.50})(Ga_{0.33}In_{0.33}Sn_{0.33})O_{3-\delta}$.

Table 2. Lattice parameter of cubic perovskite phase and reliability factors of Rietveld refinement for $(Ba_{0.50}Sr_{0.50})(Ti_{0.33}Zr_{0.33}Hf_{0.33})O_3$, $(Ba_{0.50}Sr_{0.50})(Ti_{0.33}Zr_{0.33}Sn_{0.33})O_3$, and $(Ba_{0.50}Sr_{0.50})(Ga_{0.33}In_{0.33}Sn_{0.33})O_{3-\delta}$.

| Composition | Space group | Lattice parameter [Å] | Rietveld refinement | | |
|---|---|---|---|---|---|
| | | | $R_{wp}$ (%) | $R_p$ (%) | $\chi^2$ |
| $(Ba_{0.50}Sr_{0.50})(Ti_{0.33}Zr_{0.33}Hf_{0.33})O_3$ | $Pm\bar{3}m$ | $a$ = 4.09029(2) | 4.62 | 3.59 | 4.15 |
| $(Ba_{0.50}Sr_{0.50})(Ti_{0.33}Zr_{0.33}Sn_{0.33})O_3$ | $Pm\bar{3}m$ | $a$ = 4.06732(3) | 7.11 | 5.05 | 8.65 |
| $(Ba_{0.50}Sr_{0.50})(Ga_{0.33}In_{0.33}Sn_{0.33})O_{3-\delta}$ | $Pm\bar{3}m$ | $a$ = 4.09310(1) | 7.06 | 4.85 | 6.66 |

Fig. 2 illustrates the distribution of elements in (a, b)$(Ba_{0.50}Sr_{0.50})(Ti_{0.33}Zr_{0.33}Hf_{0.33})O_3$, (c, d) $(Ba_{0.50}Sr_{0.50})(Ti_{0.33}Zr_{0.33}Sn_{0.33})O_3$, and (e, f) $(Ba_{0.50}Sr_{0.50})(Ga_{0.33}In_{0.33}Sn_{0.33})O_{3-\delta}$ achieved using EDS in (a, c, e) SEM and (b, d, f) STEM. Homogeneous distributions of all constituent elements can be clearly observed at both the micrometer and nanometer levels. Such homogeneity is due to the repeated use of HPT and calcination, as reported in some high-entropy catalytic systems [27-31].

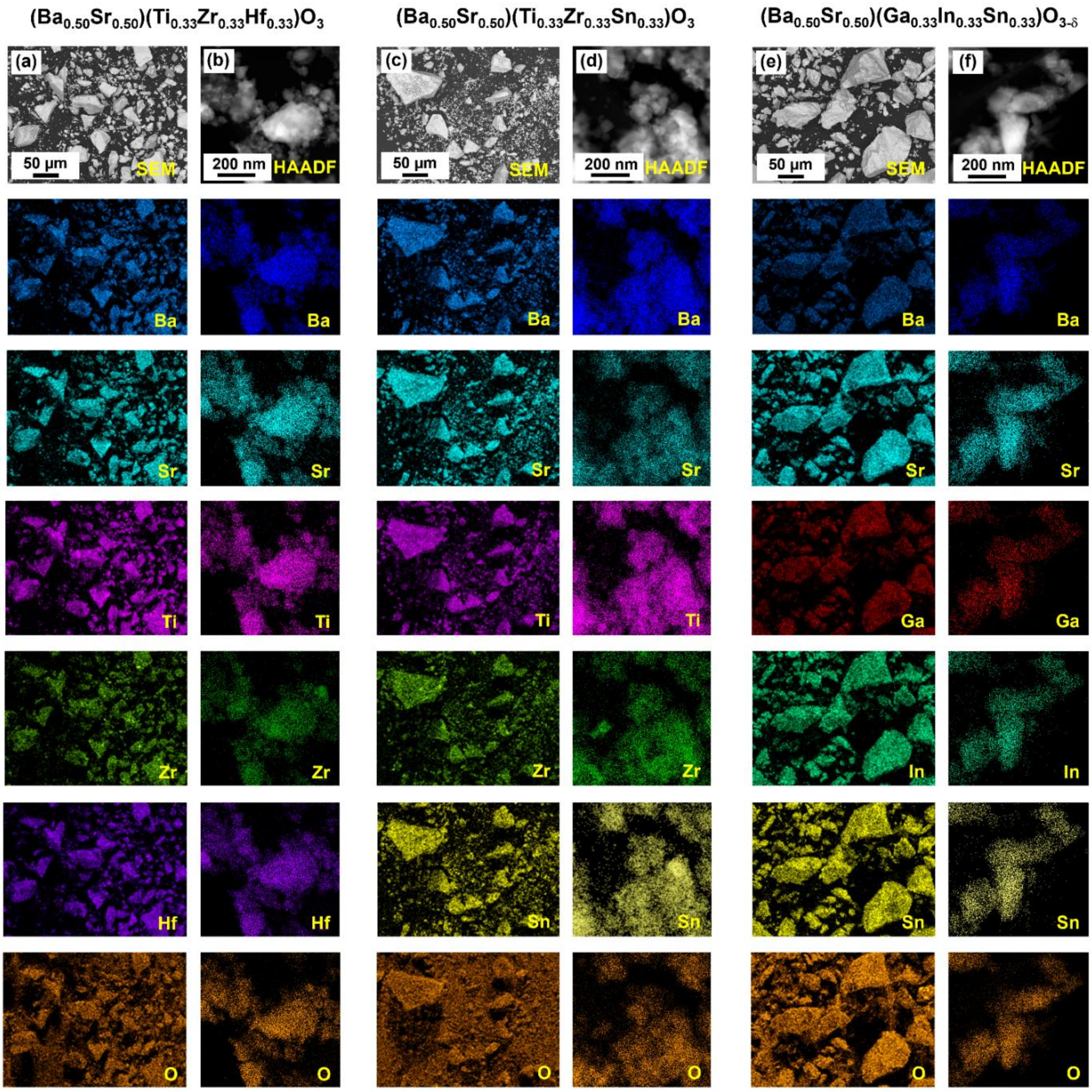

Fig. 2. Elemental homogeneity in high-entropy perovskites in micro/nanometer sizes. SEM images (left side) and HAADF micrographs (right side) with corresponding EDS mappings for (a, b) $(Ba_{0.50}Sr_{0.50})(Ti_{0.33}Zr_{0.33}Hf_{0.33})O_3$, (c, d) $(Ba_{0.50}Sr_{0.50})(Ti_{0.33}Zr_{0.33}Sn_{0.33})O_3$, and (e, f) $(Ba_{0.50}Sr_{0.50})(Ga_{0.33}In_{0.33}Sn_{0.33})O_{3-\delta}$.

XPS plots and peak deconvolution (Fig. 3a) reveal the oxygen-reduction state of the three perovskites. The spectra exhibit a dominant peak assigned to lattice O 1s, while another peak appears at higher binding energies, which should correspond to the hydroxyl group and oxygen vacancies [39]. The presence of these two peaks confirms the presence of oxygen vacancies in all three perovskites. The lattice peaks appear at 529.3 eV, 529.5 eV, and 530.8 eV for $(Ba_{0.50}Sr_{0.50})(Ti_{0.33}Zr_{0.33}Hf_{0.33})O_3$, $(Ba_{0.50}Sr_{0.50})(Ti_{0.33}Zr_{0.33}Sn_{0.33})O_3$, and $(Ba_{0.50}Sr_{0.50})(Ga_{0.33}In_{0.33}Sn_{0.33})O_{3-\delta}$, respectively. The replacement of Hf with Sn naturally increases the binding energy of O 1s due to a higher electronegativity of Sn compared to Hf. The further increase of energy by about 1.5 eV for $(Ba_{0.50}Sr_{0.50})(Ga_{0.33}In_{0.33}Sn_{0.33})O_{3-\delta}$ should originate not only from the high electronegativity of Sn and In, but also from the valence of three for In and Ga, which is lower than four for other elements in the B site. Such a valence difference generates oxygen deficiency and intrinsic vacancies in only the structure of $(Ba_{0.50}Sr_{0.50})(Ga_{0.33}In_{0.33}Sn_{0.33})O_{3-\delta}$, but the synthesis process by HPT contributes to the extrinsic vacancies in the three perovskites [6]. While these XPS data qualitatively indicate the presence of oxygen vacancies, they cannot be used in a reliable way to quantify the oxygen vacancies due to fundamental limits of XPS in analyzing only the surface, which can also be affected by adsorption of other atoms and hydroxyl groups.

Fig. 3b presents the ESR spectra of the three materials to verify the presence of vacancy defects via unpaired electrons. A dual peak with a turning point at $g = 2.008$ confirms the existence of oxygen vacancies in all three perovskites [40]. The rather low intensity of ESR peaks suggests a low fraction of vacancies with unpaired electrons, while a large fraction of vacancies with paired electrons can exist in perovskites. The presence of oxygen vacancies also suggests a potential contribution to ion conduction, demonstrating that these perovskites meet the requirements of electrolytes for the fabrication of SOFCs [7,13].

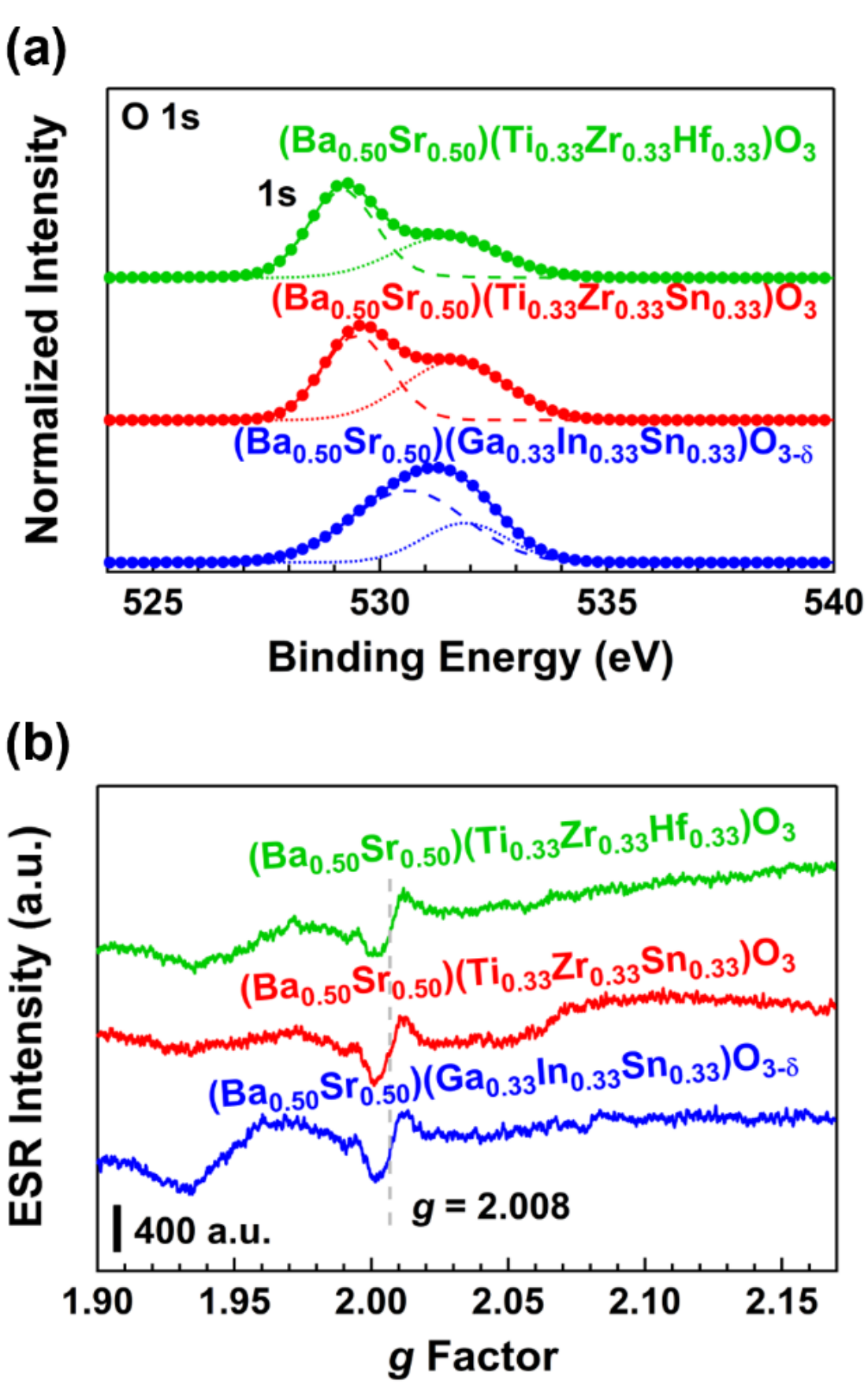

Fig. 3. Oxygen vacancies in three high-entropy perovskites. (a) XPS O 1s and (b) ESR spectra of $(Ba_{0.50}Sr_{0.50})(Ti_{0.33}Zr_{0.33}Hf_{0.33})O_3$, $(Ba_{0.50}Sr_{0.50})(Ti_{0.33}Zr_{0.33}Sn_{0.33})O_3$, and $(Ba_{0.50}Sr_{0.50})(Ga_{0.33}In_{0.33}Sn_{0.33})O_{3-\delta}$.

### *3.2. Local electronic structure*

Fig. 4 illustrates the XAS analysis of five metals (a, b) Ba, (c, d) Ti, (e, f) Hf, (g, h) Sn, and (i, j) Ga in the perovskites compared to their corresponding binary oxides, which comprises two regions: (a, c, e, g, i) XANES and (b, d, f, h, j) EXAFS. Although such XANES and EXAFS analyses do not provide quantitative relations between fuel cell performance and local electron structure, several key points regarding local electronic and geometric structures can still be inferred.

First, examination of the XANES region of the XAS spectra indicates that the absorption edges of Ba (Fig. 4a) and Ga (Fig. 4i) show an apparent similarity between the perovskites and their corresponding binary oxides. In contrast, the absorption edges of Ti (Fig. 4c) and Hf (Fig. 4e) in perovskites exhibit a slight shift toward lower energies compared with those of $TiO_2$ and $HfO_2$, suggesting differences in the local electronic environment. These differences in XANES profiles may be associated with lower oxidation states of these metals due to the absence of some metal-oxygen bonds [27]. The interpretation of the Sn absorption edge in the XAS spectra is more complex (Fig. 4g). The Sn $L_3$-edge in $SnO_2$ consists of two absorption peaks. However, the exact origin of these dual peaks has not yet been agreed upon. Previous studies have suggested that this phenomenon may arise from significant hybridization of the Sn 5s and 5p states and the O 2p states in the electronic configuration of Sn in $SnO_2$ ([Kr]$4d^{10}5s^{0}5p^{0}$) [41]. According to this model, during the $L_3$-edge transition from $2p^{3/2}$ to $5s^{1/2}$, the first peak is assigned to the $5s^0$ electronic configuration, whereas the second peak corresponds to the $5s^1L$ configuration, where L denotes a ligand-related electron-deficient state (ligand hole) [41]. Based on this interpretation, the XANES features of Sn in $SnO_2$ and in the HEPs can be compared. At the first peak, the absorption edge of Sn in $SnO_2$ appears at a higher energy than that in the HEPs, indicating differences in the local electronic configuration of Sn, possibly related to different electron density distributions between the two material systems. Conversely, at the second peak, the absorption edge of Sn in $SnO_2$ is lower than that in the HEPs. This trend may reflect enhanced Sn–O hybridization in the perovskites, leading to modifications in the binding energy and the oxidation characteristics of Sn within this complex structural environment.

Second, in the XANES region, the white line appears as a sharp absorption peak immediately after the absorption edge, reflecting the probability of electronic transitions from core levels to unoccupied states [42]. The intensity of the Ba white line (Fig. 4a) increases progressively in the order of $BaO_2$, $(Ba_{0.50}Sr_{0.50})(Ga_{0.33}In_{0.33}Sn_{0.33})O_{3-\delta}$, $(Ba_{0.50}Sr_{0.50})(Ti_{0.33}Zr_{0.33}Hf_{0.33})O_3$ and $(Ba_{0.50}Sr_{0.50})(Ti_{0.33}Zr_{0.33}Sn_{0.33})O_3$. This trend indicates a higher density of unoccupied d states for Ba in perovskites than in the corresponding oxide, likely due to variations in the local coordination environment and enhanced Ba–O hybridization in the high-entropy structure, rather than a change in the Ba valence state. For Ti (Fig. 4c), the white line intensity indicates that $(Ba_{0.50}Sr_{0.50})(Ti_{0.33}Zr_{0.33}Hf_{0.33})O_3$ and $(Ba_{0.50}Sr_{0.50})(Ti_{0.33}Zr_{0.33}Sn_{0.33})O_3$ have comparable unoccupied d states, both of which are higher than that in $TiO_2$. In contrast, Hf (Fig. 4e) and Ga (Fig. 4i) show lower white line intensities in the perovskites than in their respective oxides. For Sn (Fig. 4g), the unoccupied d-state density increases progressively from $(Ba_{0.50}Sr_{0.50})(Ti_{0.33}Zr_{0.33}Sn_{0.33})O_3$ to $SnO_2$ and then to $(Ba_{0.50}Sr_{0.50})(Ga_{0.33}In_{0.33}Sn_{0.33})O_{3-\delta}$. These opposite trends in $s^0/d^0$ and $d^{10}$ cations, with unoccupied d states, lead to localized electronic heterogeneity that hinders electron transfer [28-31]. The most extreme difference between the behaviour of $s^0/d^0$ and $d^{10}$ cations is achieved for $(Ba_{0.50}Sr_{0.50})(Ti_{0.33}Zr_{0.33}Sn_{0.33})O_3$.

Third, the Fourier-transformed EXAFS spectra in $R$ space describe the local coordination environment of different cations in the perovskites and their corresponding oxides. The first intense peak can be assigned to metal–oxygen bonds, while the subsequent peaks typically represent metal–metal interactions within the material [43]. Variations in the peak position reflect changes in interatomic distances, whereas changes in peak amplitude are related to coordination number and local structural disorder. EXAFS spectra reveal that the Ba–O bond length (Fig. 4b) increases progressively in the order $(Ba_{0.50}Sr_{0.50})(Ga_{0.33}In_{0.33}Sn_{0.33})O_{3-\delta}$, $BaO_2$, $(Ba_{0.50}Sr_{0.50})(Ti_{0.33}Zr_{0.33}Sn_{0.33})O_3$, and $(Ba_{0.50}Sr_{0.50})(Ti_{0.33}Zr_{0.33}Hf_{0.33})O_3$. The Ti–O bond length (Fig. 4d) increases from $(Ba_{0.50}Sr_{0.50})(Ti_{0.33}Zr_{0.33}Sn_{0.33})O_3$ to $(Ba_{0.50}Sr_{0.50})(Ti_{0.33}Zr_{0.33}Hf_{0.33})O_3$, and then to $TiO_2$. The Sn–O bond length (Fig. 4h) increases in the sequence $(Ba_{0.50}Sr_{0.50})(Ga_{0.33}In_{0.33}Sn_{0.33})O_{3-\delta}$, $(Ba_{0.50}Sr_{0.50})(Ti_{0.33}Zr_{0.33}Sn_{0.33})O_3$ and $SnO_2$. Conversely, the Hf–O (Fig. 4f) and Ga–O (Fig. 4j) bond lengths in the perovskites are shorter than those in their respective binary oxides. Shorter metal-oxygen bond lengths enhance proton mobility but negatively affect proton trapping, so an optimized bond length is favorable. XANES analyses also suggest that several metal–metal bonds in the perovskites differ from those in the initial oxides. This is observed by changes in radial distances, such as the replacement of Ba–Ba, Ti–Ti, and Sn–Sn bonds with Ba–metal, Ti–metal, and Sn–metal bonds, respectively. In addition, some metal–metal bonds, like Hf–Hf and Ga–Ga, are preserved. Still, their bond lengths are adjusted in the perovskites relative to those of the original oxides to maintain the structural stability of the high-entropy framework.

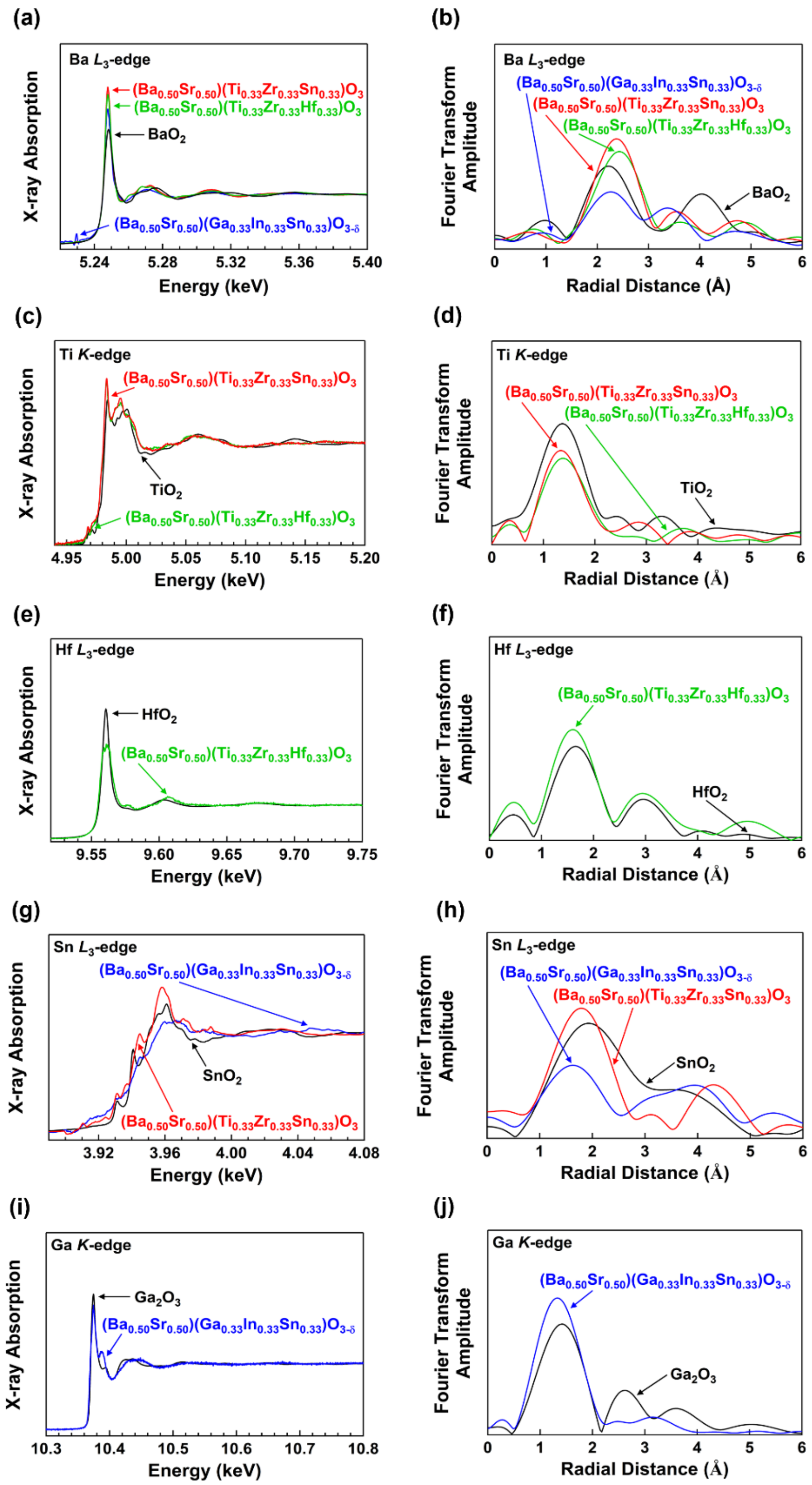

Fig. 4. Distorted local electronic structure in the vicinity of cations in high-entropy perovskites. XAS (left side) and EXAFS (right side) graphs for (a, b) Ba $L_3$, (c, d) Ti $K$, (e, f) Hf $L_3$, (g, h) Sn $L_3$, (i, j) Ga $K$ edges of $(Ba_{0.50}Sr_{0.50})(Ti_{0.33}Zr_{0.33}Hf_{0.33})O_3$, $(Ba_{0.50}Sr_{0.50})(Ti_{0.33}Zr_{0.33}Sn_{0.33})O_3$ and $(Ba_{0.50}Sr_{0.50})(Ga_{0.33}In_{0.33}Sn_{0.33})O_{3-\delta}$, and their corresponding binary oxides.

### *3.3. Electrochemical performance of fuel cells*

The cross-sectional SEM micrographs of fuel cells are presented in Fig. 5 for the cells fabricated using (a) $(Ba_{0.50}Sr_{0.50})(Ti_{0.33}Zr_{0.33}Hf_{0.33})O_3$, (b) $(Ba_{0.50}Sr_{0.50})(Ti_{0.33}Zr_{0.33}Sn_{0.33})O_3$, and (c) $(Ba_{0.50}Sr_{0.50})(Ga_{0.33}In_{0.33}Sn_{0.33})O_{3-\delta}$. These micrographs reveal distinct layers of anode, electrolyte, and cathode within the cells. It can be noted that neither an interfacial reaction layer nor severe delamination was observed in the cross-sectional SEM images, suggesting acceptable compatibility among the cell components under the present fabrication conditions. The electrolyte layers exhibit good sealing properties and clean electrolyte–electrode interfaces. After two screen-printing passes, the electrolyte layers of $(Ba_{0.50}Sr_{0.50})(Ti_{0.33}Zr_{0.33}Hf_{0.33})O_3$ (7.4 μm thick) and $(Ba_{0.50}Sr_{0.50})(Ti_{0.33}Zr_{0.33}Sn_{0.33})O_3$ (5.8 μm thick), became relatively dense, resulting in better cell than those obtained after three screen-printing passes, as evidenced by the cross-sectional morphologies (Fig. S5) and the corresponding electrochemical performance (Fig. S6). However, fabrication and sintering at 1623 K led to the formation of severe voids in $(Ba_{0.50}Sr_{0.50})(Ga_{0.33}In_{0.33}Sn_{0.33})O_{3-\delta}$. This behavior may be related to elemental volatilization, insufficient densification, and defect-induced microstructural instability during high-temperature sintering. To minimize void formation and obtain a sufficiently dense electrolyte layer (9.8 μm thick), the sintering temperature was optimized to 1573 K for 5 h. Nevertheless, even at the optimized temperature, void formation could not be completely avoided in this electrolyte.

Fig. 5d illustrates the open-circuit-voltage profiles of the three cells after the anode reduction process. After 3 hours, the open-circuit voltage of the cells containing $(Ba_{0.50}Sr_{0.50})(Ti_{0.33}Zr_{0.33}Hf_{0.33})O_3$ and $(Ba_{0.50}Sr_{0.50})(Ti_{0.33}Zr_{0.33}Sn_{0.33})O_3$ at 873 K reaches 1.13 V and 1.17 V, respectively. These values are close to the theoretical value of 1.14 V, calculated by considering the Nernst equation at 873 K under the actual operating conditions (3% $H_2O$ + $H_2$ at the anode and 3% $H_2O$ + air at the cathode). Fig. S7 presents the temperature dependence of the ionic transference number ($t_i$), calculated as the ratio of the experimentally measured to the theoretical open-circuit voltage using the Nernst equation under the actual operating conditions. The $t_i$ average values are approximately 0.96 and decrease gradually as temperature increases. The high $t_i$ values indicate minor electronic leakage at elevated temperatures [44-46]. The open-circuit voltage for the cell containing $(Ba_{0.50}Sr_{0.50})(Ga_{0.33}In_{0.33}Sn_{0.33})O_{3-\delta}$ at 873 K reaches 0.75 V, which is considerably lower than that of the two other materials. The presence of voids in the electrolyte layer should be responsible for the limited electrochemical performance of the cell containing this electrolyte.

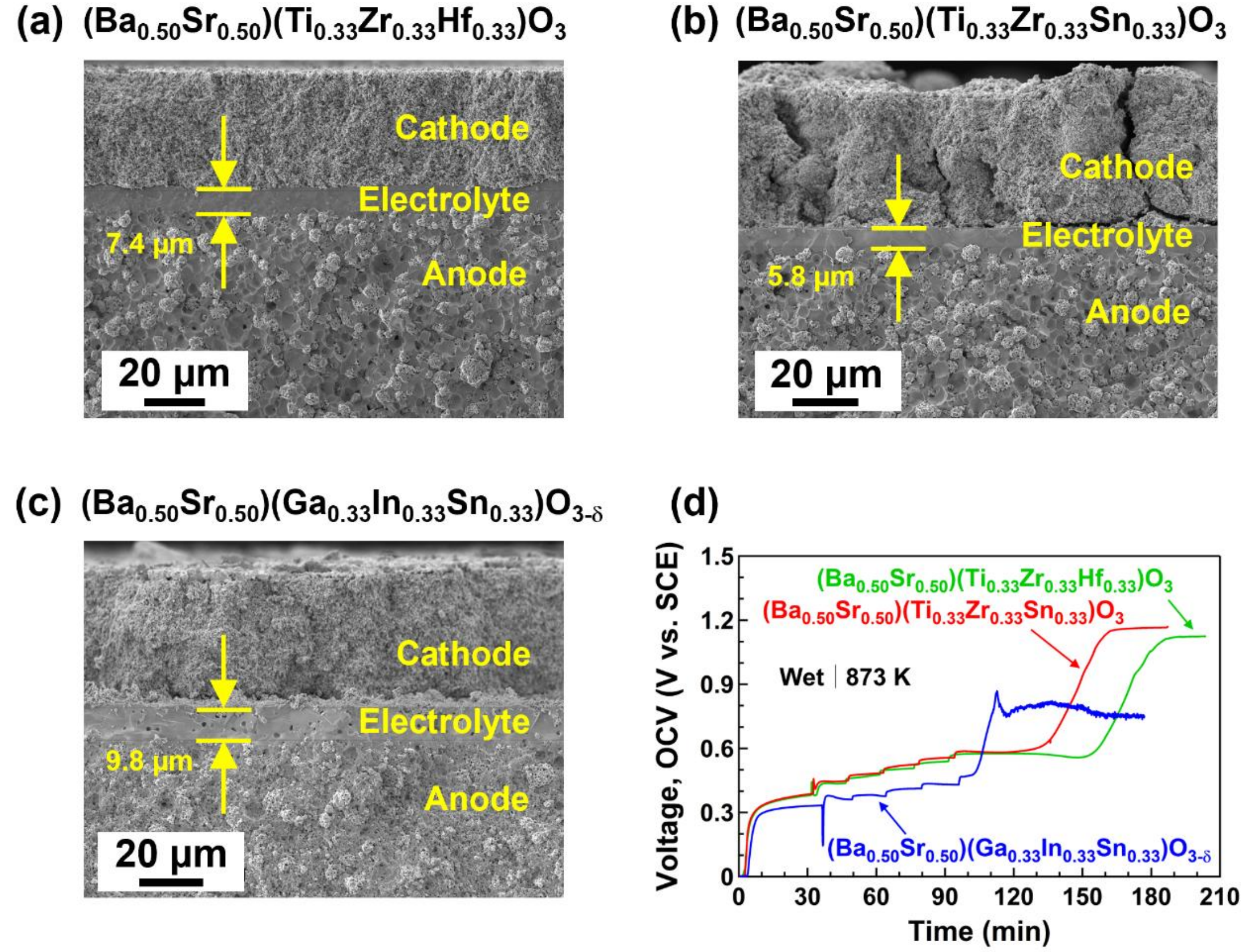

Fig. 5. Dense fuel cells with high open-circuit voltage using high-entropy perovskites as electrolytes. Cross-sectional SEM micrographs of cells fabricated using (a) $(Ba_{0.50}Sr_{0.50})(Ti_{0.33}Zr_{0.33}Hf_{0.33})O_3$, (b) $(Ba_{0.50}Sr_{0.50})(Ti_{0.33}Zr_{0.33}Sn_{0.33})O_3$, and (c) $(Ba_{0.50}Sr_{0.50})(Ga_{0.33}In_{0.33}Sn_{0.33})O_{3-\delta}$ electrolytes and (d) corresponding open-circuit voltage profiles. The micrograph in (a) was taken by back-scattered electrons, and micrographs in (b) and (c) were taken by secondary electrons.

Fig. 6 presents the EIS spectra of the fuel cells measured under different operating conditions. The spectra were fitted using the equivalent circuit $L(R_o)(R_{HF}Q_{HF})(R_{MF}Q_{MF})(R_{LF}Q_{LF})$ for $(Ba_{0.50}Sr_{0.50})(Ti_{0.33}Zr_{0.33}Hf_{0.33})O_3$ and $(Ba_{0.50}Sr_{0.50})(Ti_{0.33}Zr_{0.33}Sn_{0.33})O_3$, and $(R_o)(R_{HF}Q_{HF})(R_{MF}Q_{MF})(R_{LF}Q_{LF})$ for $(Ba_{0.50}Sr_{0.50})(Ga_{0.33}In_{0.33}Sn_{0.33})O_{3-\delta}$, where $L$, $R$, $Q$, HF, MF, and LF represent the inductance, resistance, constant phase element, high-frequency, medium-frequency, and low-frequency, respectively. Considering the influence of water vapor pressure on the overall performance of the cells [8,11], fuel cell operation was first investigated at 873 K under dry and wet gas-feed conditions. The EIS results show that under dry conditions, the ohmic resistances ($R_{ohm}$) of $(Ba_{0.50}Sr_{0.50})(Ti_{0.33}Zr_{0.33}Hf_{0.33})O_3$ and $(Ba_{0.50}Sr_{0.50})(Ti_{0.33}Zr_{0.33}Sn_{0.33})O_3$ are comparable, with a value of approximately 1.10 Ω.cm$^2$, despite the electrolyte layer of $(Ba_{0.50}Sr_{0.50})(Ti_{0.33}Zr_{0.33}Sn_{0.33})O_3$ (Fig. 5b) being thinner than that of $(Ba_{0.50}Sr_{0.50})(Ti_{0.33}Zr_{0.33}Hf_{0.33})O_3$ (Fig. 5a). However, a notable difference was observed under wet air/$H_2$ conditions, which are more representative of practical fuel-cell operation. The fuel cell based on $(Ba_{0.50}Sr_{0.50})(Ti_{0.33}Zr_{0.33}Sn_{0.33})O_3$ exhibited the lowest $R_{ohm}$ of 0.75 Ω.cm$^2$, compared with 0.90 Ω.cm$^2$ for the fuel cell based on $(Ba_{0.50}Sr_{0.50})(Ti_{0.33}Zr_{0.33}Hf_{0.33})O_3$. The results indicate that the lower resistance of the $(Ba_{0.50}Sr_{0.50})(Ti_{0.33}Zr_{0.33}Sn_{0.33})O_3$-based fuel cell is primarily attributable to intrinsic electrolyte resistance rather than to reduced electrolyte thickness achieved during fabrication and electrode-related processes. In contrast, the fuel cell based on $(Ba_{0.50}Sr_{0.50})(Ga_{0.33}In_{0.33}Sn_{0.33})O_{3-\delta}$ (Fig. 6g) exhibits nearly unchanged behavior under both dry and wet conditions. Furthermore, the high-frequency contribution in its impedance spectrum was substantially larger, indicating that electrolyte-related resistance dominates the overall electrochemical response. This observation is consistent with the presence of pores and structural defects in this electrolyte, with a thickness of approximately 10 μm.

Further electrochemical analyses were conducted under wet conditions and at different temperatures because mass transport is thermally activated, thereby strongly influencing the electrochemical performance of solid fuel cells [7,20,47]. Fig. 6 shows the EIS characteristics of the cells fabricated by (b) $(Ba_{0.50}Sr_{0.50})(Ti_{0.33}Zr_{0.33}Hf_{0.33})O_3$, (e) $(Ba_{0.50}Sr_{0.50})(Ti_{0.33}Zr_{0.33}Sn_{0.33})O_3$, and (h) $(Ba_{0.50}Sr_{0.50})(Ga_{0.33}In_{0.33}Sn_{0.33})O_{3-\delta}$ at a temperature range of 773-973 K under wet hydrogen and air supply ($P_{H_2O}$ = 1.9 kPa). It is evident that the three cells perform better at higher temperatures, with the smallest impedance spectra for all three solid cells obtained at 973 K. Notably, the $(Ba_{0.50}Sr_{0.50})(Ti_{0.33}Zr_{0.33}Sn_{0.33})O_3$ electrolyte-based cell exhibits an $R_{ohm}$ of 0.5 Ω.cm$^2$ at 973 K, which is lower than the other fuel cells: 0.65 Ω.cm$^2$ for $(Ba_{0.50}Sr_{0.50})(Ti_{0.33}Zr_{0.33}Hf_{0.33})O_3$ and 2.0 Ω.cm$^2$ for $(Ba_{0.50}Sr_{0.50})(Ga_{0.33}In_{0.33}Sn_{0.33})O_{3-\delta}$. Based on the obtained ohmic resistance values, the Arrhenius equation was applied according to Eq. 4, and the corresponding plots are presented in Fig. 6c, 6f, and 6i for $(Ba_{0.50}Sr_{0.50})(Ti_{0.33}Zr_{0.33}Hf_{0.33})O_3$, $(Ba_{0.50}Sr_{0.50})(Ti_{0.33}Zr_{0.33}Sn_{0.33})O_3$, and $(Ba_{0.50}Sr_{0.50})(Ga_{0.33}In_{0.33}Sn_{0.33})O_{3-\delta}$, respectively, to describe the temperature dependence of the ohmic resistance [48].

$$R = R_0 \exp\left(\frac{E_A}{k_B T}\right) \quad \text{or} \quad \ln R = \ln(R_0) + \frac{E_A}{1000 k_B}\frac{1000}{T} \tag{4}$$

where $R$ is the ohmic area-specific resistance at absolute temperature $T$, $R_0$ is the pre-exponential factor, $k_B$ is the Boltzmann constant, and $E_A$ is the activation energy for charge transport. It is observed that all three fuel cells exhibit activation energies of approximately 0.3 eV. This value is consistent with proton migration in oxide lattices and is considerably lower than the activation energies typically reported for oxygen ion conduction [49-51]. Taken with the evidence of oxygen vacancies revealed by XPS and ESR analyses, these results suggest that proton transport makes a significant contribution to the overall conductivity of these fuel cells.

To further clarify the electrochemical performance of HEPs, Fig. S8a, b, and c describe the impedance spectra of the fuel cells based on $(Ba_{0.50}Sr_{0.50})(Ti_{0.33}Zr_{0.33}Hf_{0.33})O_3$, $(Ba_{0.50}Sr_{0.50})(Ti_{0.33}Zr_{0.33}Sn_{0.33})O_3$, and $(Ba_{0.50}Sr_{0.50})(Ga_{0.33}In_{0.33}Sn_{0.33})O_{3-\delta}$, respectively. The measurements were conducted at 873 K under wet air/$H_2$ conditions, and deconvoluted into individual resistive contributions using the distribution of relaxation times (DRT) analysis. The resulting DRT spectra can be separated into three frequency regions: high-frequency ($10^3$–$10^5$ Hz), medium-frequency ($10^1$–$10^3$ Hz), and low-frequency ($10^{-1}$–$10^1$ Hz). The DRT analysis parameters are summarized in Table S1. For the $(Ba_{0.50}Sr_{0.50})(Ga_{0.33}In_{0.33}Sn_{0.33})O_{3-\delta}$-based fuel cell, the electrolyte was not fully dense, resulting in poor electrochemical performance and a relatively high fitting error. In a $(Ba_{0.50}Sr_{0.50})(Ti_{0.33}Zr_{0.33}Hf_{0.33})O_3$-based fuel cell, the mid-frequency process dominates the impedance response and is likely responsible for overall resistance. In contrast, the $(Ba_{0.50}Sr_{0.50})(Ti_{0.33}Zr_{0.33}Sn_{0.33})O_3$ and $(Ba_{0.50}Sr_{0.50})(Ga_{0.33}In_{0.33}Sn_{0.33})O_{3-\delta}$-based fuel cells exhibit a relatively large contribution from the high-frequency process and a small contribution from the mid-frequency process, indicating a more balanced electrochemical response. These distinct DRT characteristics indicate that the electrochemical processes governing the $(Ba_{0.50}Sr_{0.50})(Ti_{0.33}Zr_{0.33}Hf_{0.33})O_3$-based fuel cell differ from those operating in the other two fuel cells.

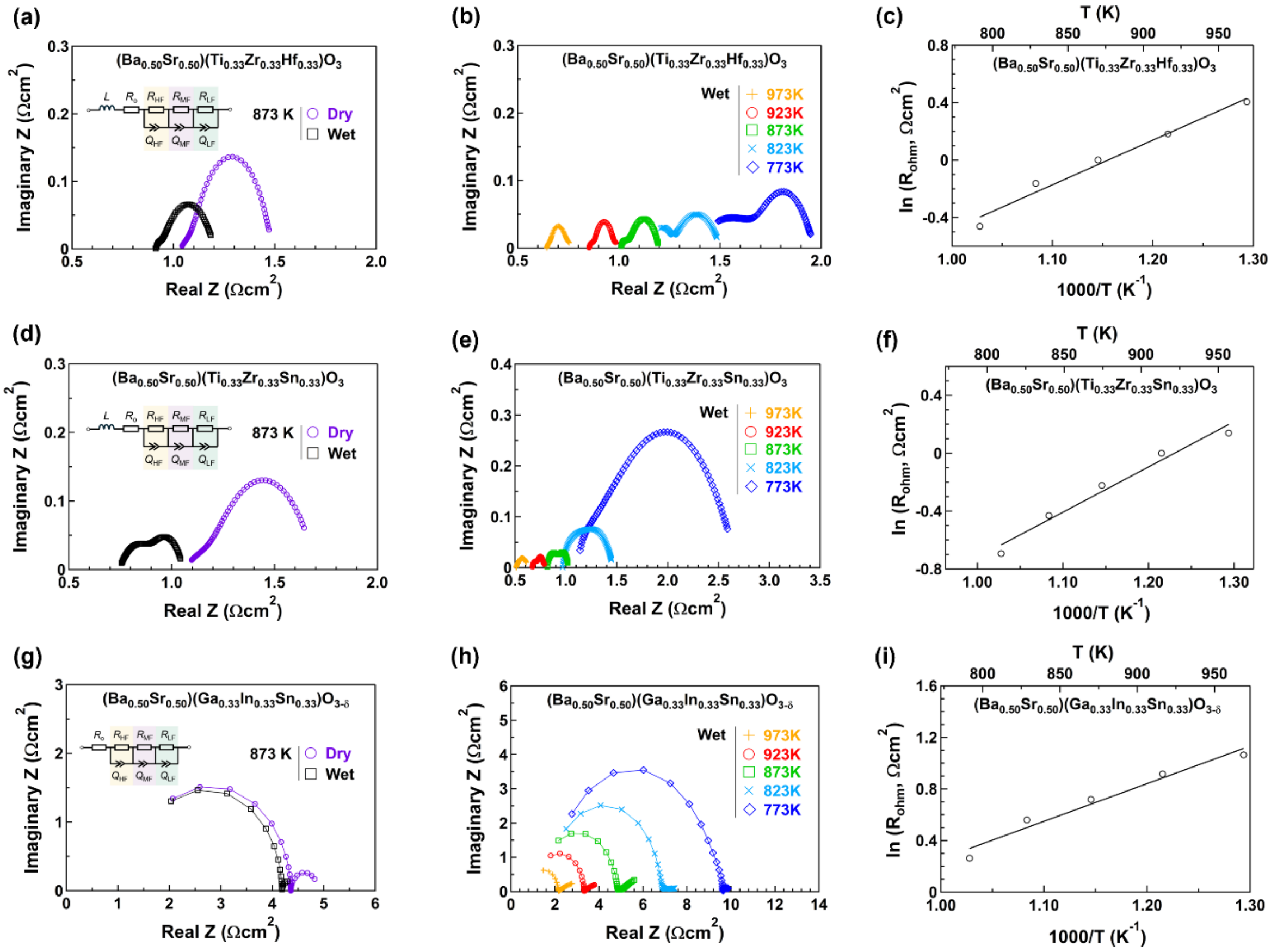

Fig. 6. Low ohmic resistance under wet conditions and at high temperature for fuel cells fabricated using high-entropy perovskites as electrolytes. EIS spectra of fuel cells with (a, b) $(Ba_{0.50}Sr_{0.50})(Ti_{0.33}Zr_{0.33}Hf_{0.33})O_3$, (d, e) $(Ba_{0.50}Sr_{0.50})(Ti_{0.33}Zr_{0.33}Sn_{0.33})O_3$ and (g, h) $(Ba_{0.50}Sr_{0.50})(Ga_{0.33}In_{0.33}Sn_{0.33})O_{3-\delta}$ electrolytes achieved (a, d, g) at 873 K under dry and wet conditions, and (b, e, h) at different temperatures under wet condition. Arrhenius plot of ohmic resistance and activation energy for (c) $(Ba_{0.50}Sr_{0.50})(Ti_{0.33}Zr_{0.33}Hf_{0.33})O_3$, (f) $(Ba_{0.50}Sr_{0.50})(Ti_{0.33}Zr_{0.33}Sn_{0.33})O_3$, and (i) $(Ba_{0.50}Sr_{0.50})(Ga_{0.33}In_{0.33}Sn_{0.33})O_{3-\delta}$ electrolyte-based fuel cells.

To further clarify the electrochemical performance of HEPs, current-voltage characteristics of three cells were measured under wet conditions over the temperature range of 773-973 K. As shown in Fig. 7 for cells fabricated using (a) $(Ba_{0.50}Sr_{0.50})(Ti_{0.33}Zr_{0.33}Hf_{0.33})O_3$, (b) $(Ba_{0.50}Sr_{0.50})(Ti_{0.33}Zr_{0.33}Sn_{0.33})O_3$, and (c) $(Ba_{0.50}Sr_{0.50})(Ga_{0.33}In_{0.33}Sn_{0.33})O_{3-\delta}$, the current density increases with rising operating temperatures for the three cells. The peak power density described in Fig.7b, d, and f of the fuel cells with $(Ba_{0.50}Sr_{0.50})(Ti_{0.33}Zr_{0.33}Hf_{0.33})O_3$, $(Ba_{0.50}Sr_{0.50})(Ti_{0.33}Zr_{0.33}Sn_{0.33})O_3$, and $(Ba_{0.50}Sr_{0.50})(Ga_{0.33}In_{0.33}Sn_{0.33})O_{3-\delta}$ electrolytes reaches 0.42, 0.53, and 0.06 W.cm$^{-2}$, respectively. These results, together with EIS analyses, suggest that, while these perovskites function as electrolytes, $(Ba_{0.50}Sr_{0.50})(Ti_{0.33}Zr_{0.33}Sn_{0.33})O_3$ exhibits the best electrochemical performance. It should be noted that the good performance of $(Ba_{0.50}Sr_{0.50})(Ti_{0.33}Zr_{0.33}Sn_{0.33})O_3$ is not because of its low electrolyte thickness, as its thicker electrolyte (10.6 μm thick) also shows better performance than the two other perovskites (Fig. S6).

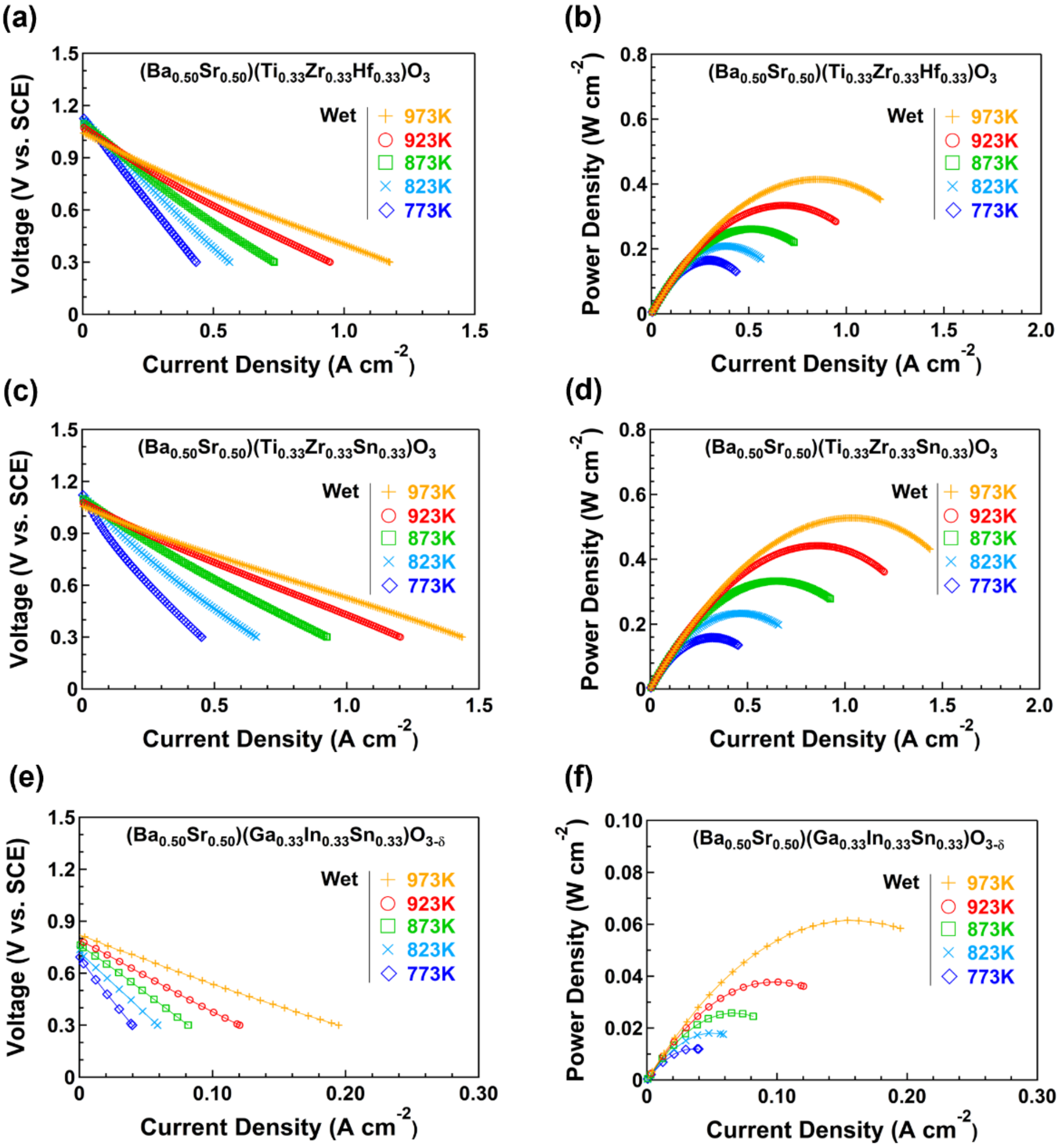

Fig. 7. High peak power density of fuel cells fabricated using high-entropy perovskites as electrolytes. Current-voltage curves (a, c, e) and polarization curve (b, d, f) of fuel cells with (a, b) $(Ba_{0.50}Sr_{0.50})(Ti_{0.33}Zr_{0.33}Hf_{0.33})O_3$, (c, d) $(Ba_{0.50}Sr_{0.50})(Ti_{0.33}Zr_{0.33}Sn_{0.33})O_3$, and (e, f) $(Ba_{0.50}Sr_{0.50})(Ga_{0.33}In_{0.33}Sn_{0.33})O_{3-\delta}$ electrolytes achieved at different conditions under a wet environment.

Fig. 8a presents a 40 h stability test conducted at an applied voltage of 0.9 V under wet operating conditions at 873 K for $(Ba_{0.50}Sr_{0.50})(Ti_{0.33}Zr_{0.33}Sn_{0.33})O_3$ as the best-performing electrolyte. The results demonstrate that the fuel cell exhibits a gradual reduction in performance during the testing time. Despite a reduction in performance over time, cross-sectional analysis of the fuel cell after the stability test (Fig. 8b) confirmed that the electrolyte layer remained intact, with no evidence of significant structural degradation. This stability issue needs to be further addressed in future studies by considering electrode degradation, interfacial reactions, microstructural coarsening, and gas leakage, as well as by modifying the overall fuel cell structure and fabrication.

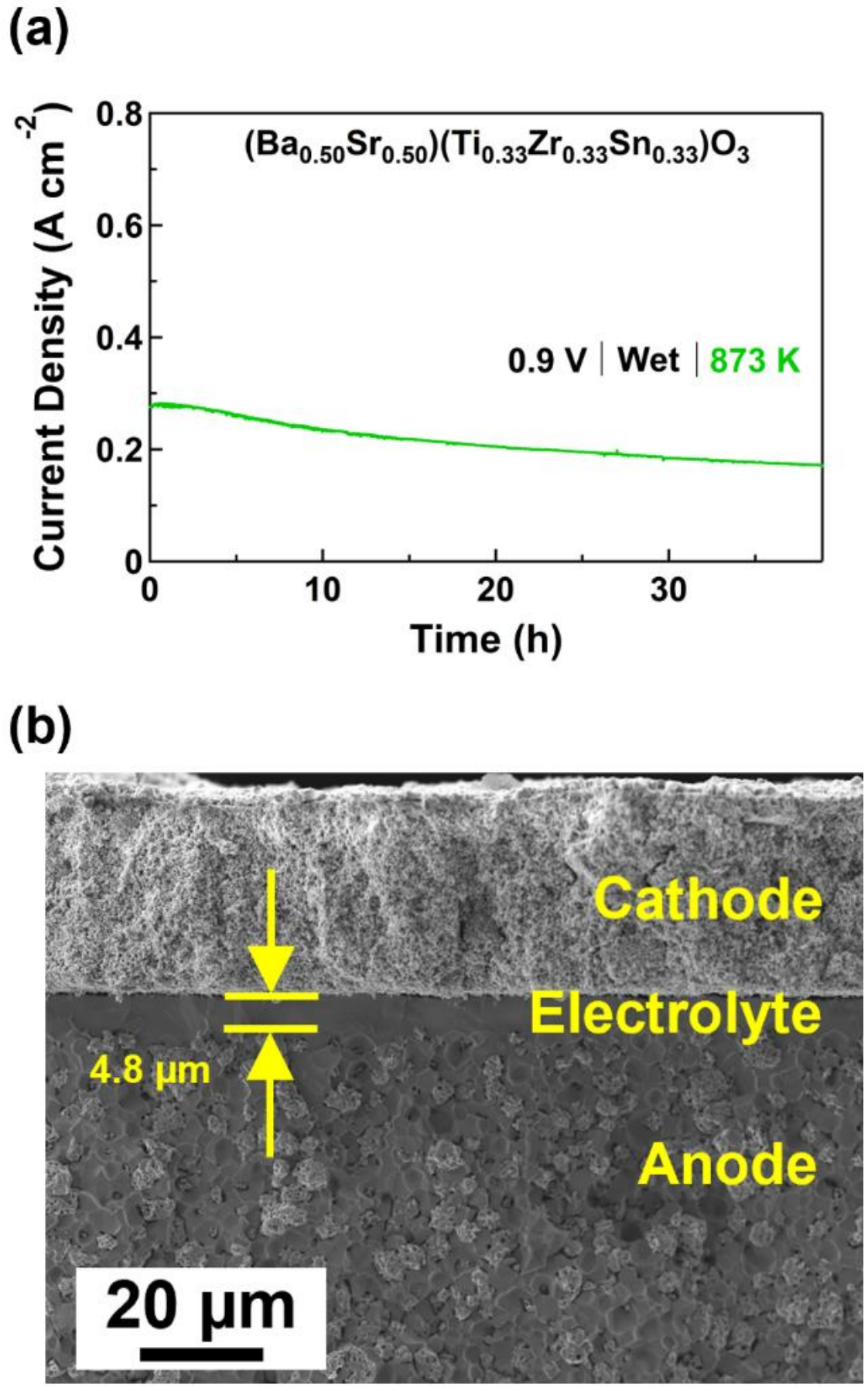

Figure 8. Moderate stability of fuel cell fabricated by high-entropy perovskite as electrolyte. (a) Stability test and (b) cross-sectional SEM micrograph, taken by secondary electrons, of a fuel cell containing $(Ba_{0.50}Sr_{0.50})(Ti_{0.33}Zr_{0.33}Sn_{0.33})O_3$ as electrolyte.

## 4. Discussion

Three rare-earth-metal-free high-entropy perovskites were synthesized in this study with the shared characteristic that the A-site cations consist of Ba and Sr with $s^0$ cationic configurations, while the B-site cations are combinations of three metals forming either $d^0$, $d^{10}$, or mixed $d^0/d^{10}$ cations. These perovskites were used as electrolytes in fuel cells, where the comparison of their electrochemical performance confirmed that the perovskite with mixed $s^0/d^0/d^{10}$ cations, $(Ba_{0.50}Sr_{0.50})(Ti_{0.33}Zr_{0.33}Sn_{0.33})O_3$, exhibits the best performance. The performance of these high-entropy perovskites can be discussed in terms of (i) crystal structure, (ii) local electronic structure and cationic configuration, and (iii) oxygen vacancies.

First, considering the crystal structure, the three perovskites have a similar crystal structure, but $(Ba_{0.50}Sr_{0.50})(Ti_{0.33}Zr_{0.33}Sn_{0.33})O_3$ has the smallest lattice parameter, which is more appropriate for ion conductivity. Moreover, the Goldschmidt tolerance factors of $(Ba_{0.50}Sr_{0.50})(Ti_{0.33}Zr_{0.33}Hf_{0.33})O_3$, $(Ba_{0.50}Sr_{0.50})(Ti_{0.33}Zr_{0.33}Sn_{0.33})O_3$, and $(Ba_{0.50}Sr_{0.50})(Ga_{0.33}In_{0.33}Sn_{0.33})O_{3-\delta}$ can be calculated as 0.946, 0.949, and 0.935, respectively [32]. Among them, $(Ba_{0.50}Sr_{0.50})(Ti_{0.33}Zr_{0.33}Sn_{0.33})O_3$ has the closest value of tolerance factor to one, which is the value for an ideal perovskite structure. Such a structure, together with the cation-size differences among B-site elements and the high configurational disorder induced by the high-entropy effect, may lead to moderate local lattice distortion. These features are considered beneficial for ion transport [51]. Therefore, the tolerance factor analysis suggests that $(Ba_{0.50}Sr_{0.50})(Ti_{0.33}Zr_{0.33}Sn_{0.33})O_3$ has a higher potential for ion conductivity.

Second, regarding the local electronic structure, the EXAFS spectra for the Ba absorption edge (Fig. 4b) show that the intensity of the first peak for $(Ba_{0.50}Sr_{0.50})(Ti_{0.33}Zr_{0.33}Sn_{0.33})O_3$ is higher than that of $(Ba_{0.50}Sr_{0.50})(Ti_{0.33}Zr_{0.33}Hf_{0.33})O_3$ and $(Ba_{0.50}Sr_{0.50})(Ga_{0.33}In_{0.33}Sn_{0.33})O_{3-\delta}$, reflecting a higher density of unoccupied d states for Ba in $(Ba_{0.50}Sr_{0.50})(Ti_{0.33}Zr_{0.33}Sn_{0.33})O_3$ compared with the two other perovskites. In addition, EXAFS analysis (Fig. 4b) also indicates that the Ba–O bond length in $(Ba_{0.50}Sr_{0.50})(Ti_{0.33}Zr_{0.33}Sn_{0.33})O_3$ is longer. Longer A–O bonds (e.g., Ba–O) are believed to facilitate the rotational motion of hydroxide ions within the perovskite lattice [52]. The rotation of hydroxide ions, coupled with proton transfer, may contribute to proton diffusion in proton-conducting electrolytes [7,52]. Regarding the role of the B-site cations, XANES spectra (Fig. 4c, 4e, 4i) show a slight decrease in the binding energy of Ti and Hf in perovskites compared with their corresponding binary oxides, while the electronic state of Ga remains nearly unchanged. For Sn (Fig. 4g), changes in the local environment are reflected by an increased degree of Sn–O hybridization in the HEPs with the lowest density of unoccupied d states for Ba in $(Ba_{0.50}Sr_{0.50})(Ti_{0.33}Zr_{0.33}Sn_{0.33})O_3$. These opposite trends of $s^0/d^0$ and $d^{10}$ cations in having unoccupied d states $(Ba_{0.50}Sr_{0.50})(Ti_{0.33}Zr_{0.33}Sn_{0.33})O_3$ should lead to localized electronic heterogeneity, hindering electron transfer [28-31]. Previous studies have also reported that Sn doping can improve electrolyte performance due to its relatively small ionic radius and high electronegativity [53,54].

Third, regarding oxygen vacancies, the O 1s XPS (Fig. 3a) and ESR spectra (Fig. 3b) for all three materials confirm their presence in the crystal lattice. In ideal perovskite electrolytes such as $BaZrO_3$ or $BaCeO_3$, intrinsic protons are typically absent [7]. Therefore, a doping strategy is commonly employed to create oxygen vacancies, which can influence the defect and transport properties of the electrolytes. In three high-entropy perovskites, $(Ba_{0.50}Sr_{0.50})(Ga_{0.33}In_{0.33}Sn_{0.33})O_{3-\delta}$ can form intrinsic oxygen vacancies due to the lower valence of In and Ga. Consequently, this material is expected to possess a higher concentration of hydration-active defect sites. However, the HEP exhibits relatively low stability and is prone to void formation at high temperatures, making it technically difficult to study the role of intrinsic vacancies. For the two other materials, oxygen vacancies are extrinsic and are formed during solid-state reaction synthesis [33]. The use of HPT during oxide synthesis can generate oxygen vacancies while maintaining some vacancies even after high-temperature sintering without doping [55,56]. Future studies should consider quantifying the role of oxygen vacancies in HEP fuel cells, incorporating intrinsic and extrinsic vacancies, and mixed $s^0/d^0/d^{10}$ cations.

Finally, it is worth comparing the performance of fuel cells fabricated using rare-earth-metal-free high-entropy perovskites as electrolytes in this study with the reported data in the literature. Table 3 compares the hydrogen-to-electricity conversion performance of SOFCs with different anodes, electrolytes, and cathodes, and at varying operating temperatures [57-62]. Although various parameters, such as the thickness of electrolyte, anode type, cathode type, optimization of sintering, humidity, gas flow rates, and material surface area, can affect such comparisons, Table 3 suggests that the fuel cell fabricated using the $(Ba_{0.50}Sr_{0.50})(Ti_{0.33}Zr_{0.33}Sn_{0.33})O_3$-based electrolyte shows reasonably high performance compared with other reported proton-conducting perovskites, despite the absence of intrinsic vacancies and rare-earth metals. These results underscore the significance of HEPs, architectured by mixed $s^0/d^0/d^{10}$ cations, as promising electrolyte materials for fuel cells. However, the $s^0/d^{10}$ HEP $(Ba_{0.50}Sr_{0.50})(Ga_{0.33}In_{0.33}Sn_{0.33})O_{3-\delta}$ is not satisfactory as an effective electrolyte and requires further optimization, particularly to avoid void formation. In addition, future studies should include more detailed investigations, such as proton conductivity measurements under humidified and dry $H_2$ atmospheres, as well as Hebb–Wagner or electromotive force measurements to determine the ionic/electronic transference numbers. Arrhenius plots and activation energy analysis under humid and dry conditions should also be performed to further clarify the proton transport behavior. Furthermore, thermogravimetric analysis or infrared spectroscopy after hydration should be conducted to verify the incorporation of $OH^-/H_2O$.

Table 3. Comparison of performance of fuel cells fabricated in this study using high-entropy perovskites as electrolytes, and of protonic ceramic fuel cells with different configurations and temperatures (PPD refers to peak power density, and OCV refers to open-circuit voltage)

| Anode | Electrolyte (thickness) | Cathode | $T_{sintering}$ (K, h) | $T_{treatment}$ (K) | OCV (V vs. SCE) | PPDs ($W.cm^{-2}$) | Ref. |
|---|---|---|---|---|---|---|---|
| $Ni/SrZr_{0.5}Ce_{0.4}Y_{0.1}O_{3-\delta}$ | $(Ba_{0.50}Sr_{0.50})(Ti_{0.33}Zr_{0.33}Hf_{0.33})O_3$ (7.4 μm) | $(Ba_{0.50}La_{0.50})CoO_{3-\delta}$ | 1673, 10 | 973 | 1.13 | 0.42 | This study |
| $Ni/SrZr_{0.5}Ce_{0.4}Y_{0.1}O_{3-\delta}$ | $(Ba_{0.50}Sr_{0.50})(Ti_{0.33}Zr_{0.33}Sn_{0.33})O_3$ (5.8 μm) | $(Ba_{0.50}La_{0.50})CoO_{3-\delta}$ | 1653, 10 | 973 | 1.16 | 0.53 | This study |
| $Ni/SrZr_{0.5}Ce_{0.4}Y_{0.1}O_{3-\delta}$ | $(Ba_{0.50}Sr_{0.50})(Ga_{0.33}In_{0.33}Sn_{0.33})O_{3-\delta}$ (9.8 μm) | $(Ba_{0.50}La_{0.50})CoO_{3-\delta}$ | 1573, 5 | 973 | 0.75 | 0.06 | This study |
| $Ni/BaSn_{0.16}Zr_{0.24}Ce_{0.35}Y_{0.1}Yb_{0.1}Dy_{0.05}O_{3-\delta}$ | $BaSn_{0.16}Zr_{0.24}Ce_{0.35}Y_{0.1}Yb_{0.1}Dy_{0.05}O_{3-\delta}$ (~ 45 μm) | $BaCo_{0.4}Fe_{0.4}Zr_{0.1}Y_{0.1}O_{3-\delta}$ | 1823, 10 | 873 | 1.02 | 0.32 | [57] |
| $Ni/BaZr_{0.1}Ce_{0.7}Y_{0.1}Yb_{0.1}O_{3-\delta}$ | $BaZr_{0.1}Ce_{0.7}Y_{0.1}Yb_{0.1}O_{3-\delta}$ (6.5 μm) | $BaCo_{0.2}Zn_{0.2}Ga_{0.2}Zr_{0.2}Y_{0.2}O_{3-\delta}$ | 1723, 5 | 873 | - | 0.29 | [58] |
| $Ni/BaCe_{0.5}Zr_{0.3}Dy_{0.2}O_{3-\delta}$ | $BaCe_{0.5}Zr_{0.3}Dy_{0.2}O_{3-\delta}$ (30 μm) | $Sr_3Fe_2O_{7-\delta}$ | 1723, 5 | 973 | 0.98 | 0.29 | [59] |
| $Ni/BaZr_{0.8}In_{0.2}O_{3-\delta}$ | $BaZr_{0.8}In_{0.2}O_{3-\delta}$ (10 – 13 μm) | $La_{0.6}Sr_{0.4}Co_{0.2}Fe_{0.8}O_{3-\delta}$ | 1723, 2 | 873 | 1.01 | 0.14 | [60] |
| $Ni/Ba(Zr_{0.1}Ce_{0.7}Y_{0.2})O_{3-\delta}$ | $Ba(Zr_{0.1}Ce_{0.7}Y_{0.2})O_{3-\delta}$ (10 μm) | $La_{0.6}Sr_{0.4}Co_{0.2}Fe_{0.8}O_{3-\delta}$ | 1723, 5 | 973 | 1.01 | 0.5 | [61] |
| $BaZr_{0.1}Ce_{0.7}Y_{0.1}Yb_{0.1}O_{3-\delta}$ | $BaZr_{0.1}Ce_{0.7}Y_{0.1}Yb_{0.1}O_{3-\delta}$ | $BaZr_{0.1}Ce_{0.7}Y_{0.1}Yb_{0.1}O_{3-\delta}$/ $SrCo_{0.4}Fe_{0.4}Zr_{0.1}Y_{0.1}O_{3-\delta}$. | 1723, 5 | 973 | 1.02 | 0.68 | [62] |

## 5. Conclusions

In this investigation, three novel electrolyte materials $(Ba_{0.50}Sr_{0.50})(Ti_{0.33}Zr_{0.33}Hf_{0.33})O_3$ with $s^0/d^0$ cations, $(Ba_{0.50}Sr_{0.50})(Ga_{0.33}In_{0.33}Sn_{0.33})O_{3-\delta}$ with $s^0/d^{10}$ cations, and $(Ba_{0.50}Sr_{0.50})(Ti_{0.33}Zr_{0.33}Sn_{0.33})O_3$ with mixed $s^0/d^0/d^{10}$ cations were integrated as electrolytes into into ion-conducting solid oxide fuel cells. The three oxides have unique electronic structures, featuring tailored unoccupied d-orbital states, favorable local metal-oxygen bond lengths, and extrinsic oxygen vacancies, with $(Ba_{0.50}Sr_{0.50})(Ti_{0.33}Zr_{0.33}Sn_{0.33})O_3$ having the most suitable structural and electronic characteristics. Notably, $(Ba_{0.50}Sr_{0.50})(Ti_{0.33}Zr_{0.33}Sn_{0.33})O_3$ exhibited a maximum power density of 0.53 $W.cm^{-2}$ at 973 K. This study introduces high-entropy perovskites with mixed $s^0/d^0/d^{10}$ cations as new rare-earth-metal-free electrolyte materials for advanced fuel cells.

**CRediT Authorship Contribution Statement**

Ho Truong Nam Hai: Conceptualization, Methodology, Analysis, Validation, Writing – review & editing. Rishad Kunafiev: Conceptualization, Methodology, Analysis, Validation, Writing – review & editing. Kwati Leonard: Conceptualization, Methodology, Validation, Review & editing. Kaveh Edalati: Conceptualization, Methodology, Validation, Review & editing.

**Declaration of Competing Interest**

The authors do not have any competing financial interests or personal relations that can influence the results presented in this manuscript.

**Data availability**

Data will be made available on request.

**Acknowledgments**

The author H.T.N.H. thanks the Yoshida Scholarship Foundation (YSF) for a Ph.D. scholarship. This study is partly supported through a grant from the ASPIRE project of the Japan Science and Technology Agency (JST) (Grant Number: JPMJAP2332).